\documentclass[aps,prb,twocolumn,groupedaddress]{revtex4-1}
\usepackage{graphics}
\usepackage{graphicx}
\usepackage{epstopdf}
\usepackage{braket}
\usepackage{bm}
\usepackage{amsmath}
\usepackage{caption}
\usepackage{subcaption}
\usepackage{caption}
\usepackage{color}
\usepackage{ulem}
\usepackage{soul}

\begin{document}

\title{ Unconventional correlated metallic behavior due to interorbital Coulomb interaction }

\author{N. Aucar Boidi}
\email[]{nair.aucar@cab.cnea.gov.ar}
\affiliation{Centro At\'{o}mico Bariloche and Instituto Balseiro, 8400 Bariloche, Argentina}

\author{A.~P. Kampf}
\affiliation{Institut f\"ur Physik, Theoretische Physik III,
Elektronische Korrelationen und Magnetismus, \\ Universit\"at Augsburg,
86135 Augsburg, Germany}

\author{K. Hallberg}
\affiliation{Instituto de Nanociencia y Nanotecnolog\'{\i}a CNEA-CONICET, Centro At\'{o}mico Bariloche and Instituto Balseiro, 8400 Bariloche, Argentina}

\date{\today}

\begin{abstract}
We study the non-degenerate one dimensional two-orbital Hubbard model with interorbital Coulomb interaction. By means of the density-matrix renormalization group technique, we calculate the local single-particle density of states and the optical conductivity at zero temperature. 
We find that a finite interorbital Coulomb repulsion $V$ generates a new class of states within the Mott-Hubbard band which has a large weight of holon-doublon pairs, which we hence call the holon-doublon band (HDB). When $V$ is sufficiently large, the HDB specifies the gapless low-energy excitations, and the system becomes an unconventional correlated metal. Optical conductivity results resolve different metallic behaviors for zero and finite interaction $V$. Compared to the case without interorbital interaction, the conductivity is strongly reduced in the correlated holon-doublon metal for finite $V$. In addition, the absorption spectrum is dominated by the HDB, which is clearly distinguishable from the Mott-Hubbard band.

\end{abstract}

\maketitle

\section{Introduction}
\label{intro}

The study of strongly correlated electron systems is one of the most challenging topics in the current research areas of condensed matter physics. Even the simplest models like the Heisenberg or the Hubbard model are still on vogue because of their connection to emergent and unconventional behaviors such as correlation-driven phase transitions or high-temperature superconductivity. 

In the past decades, in parallel to remarkable experimental progress, in particular in electron spectroscopy techniques such as angular resolved photoemission spectroscopy (ARPES)\cite{ARPES},
there have been important developments and improvements of efficient numerical techniques which have paved the way for a more detailed theoretical understanding of the properties of correlated systems.
Among these numerical techniques we highlight the Density-Matrix Renormalization Group (DMRG)\cite{White,Karen1,Uli} together with its modern versions in matrix product states and Tensor Network representations \cite{Uli2,banuls}, and the concomitant developments for calculating dynamical response functions\cite{Karen2,ramasesha,KW}, together with other methods such as the Dynamical Mean Field Theory (DMFT)\cite{dmft} using efficient impurity solvers based on the DMRG\cite{garcia,EPLreview}.

In the present work, we study the dynamical properties of the non-degenerate two-orbital Hubbard model with interorbital interaction in one-dimension at zero temperature. This model is relevant for understanding the microscopic physics in low-dimensional correlated materials with localized and delocalized electrons, e.g. iron-selenide two-leg ladder materials\cite{Takubo,Yamauchi,Takahashi,Ying, Zhang}, some of which behave in a manner compatible with orbital selective Mott physics.\cite{Caron, Rincon, Dong}.

Previous results on similar models have reported holon-doublon excitations at high energies \cite{fulde,wang,picos,karski}, which are accessible and have been examined as photon-induced states \cite{prelovsek,rinconfeiguin}. In addition, the emergence of novel structures within the correlated Hubbard bands was later observed in numerical studies for the two-orbital Kanamori-Hubbard model on a Bethe lattice using DMFT.\cite{yurielhd,yurieldop,Yashar} 
Subsequent work on the doped one-dimensional, degenerate version of this model confirmed the existence of in-gap states generated by $V$ and observed that for large dopings those excitations are formed mainly by interorbital holon-doublon (HD) states; their energies follow approximately the HD states of the atomic limit.\cite{nair} Also, additional Hund bands were reported recently\cite{dagottohund} as well as excitonic density waves and biexcitons in a highly doped strongly interacting version of the Kanamori-Hubbard model.\cite{feiguinexcitons}

By virtue of the high accuracy provided by the DMRG for static and dynamical properties, we find in the non-degenerate two-orbital model that a narrow band is formed within the Mott-Hubbard gap as a consequence of a finite interorbital Coulomb repulsion $V$. This band has a large projection onto interorbital holon-doublon excitons and constitutes the gapless low-energy excitations for sufficiently large values of $V$, thereby forming an unconventional correlated metal, as becomes evident in the optical conductivity.

\section{Model and method}

Specifically, we study the two-orbital Hubbard model in one dimension with broken orbital degeneracy and interorbital Coulomb repulsion:

\begin{equation}
\begin{split}
H=-t\sum_{j \alpha\sigma} \left( c_{j,\alpha\sigma}^{\dagger}c_{j+1,\alpha\sigma} + H.c.\right) - (V_g -\epsilon &)\sum_{j \alpha}n_{j,\alpha} \\
- \Delta\sum_{j} n_{j,2} + \sum_{j}{H}_{j}\mbox{,}
\end{split}
\label{eq:KHM}
\end{equation}
where the on-site interactions $H_j$ are:
\begin{equation}
H_j=U\sum_{\alpha}n_{j,\alpha\uparrow}n_{j,\alpha\downarrow}+V\sum_{\sigma\sigma'}n_{j,1\sigma}n_{j,2\sigma'} .
\label{eq:interaction}
\end{equation}

\begin{figure}[h!]
  \includegraphics[scale=0.33]{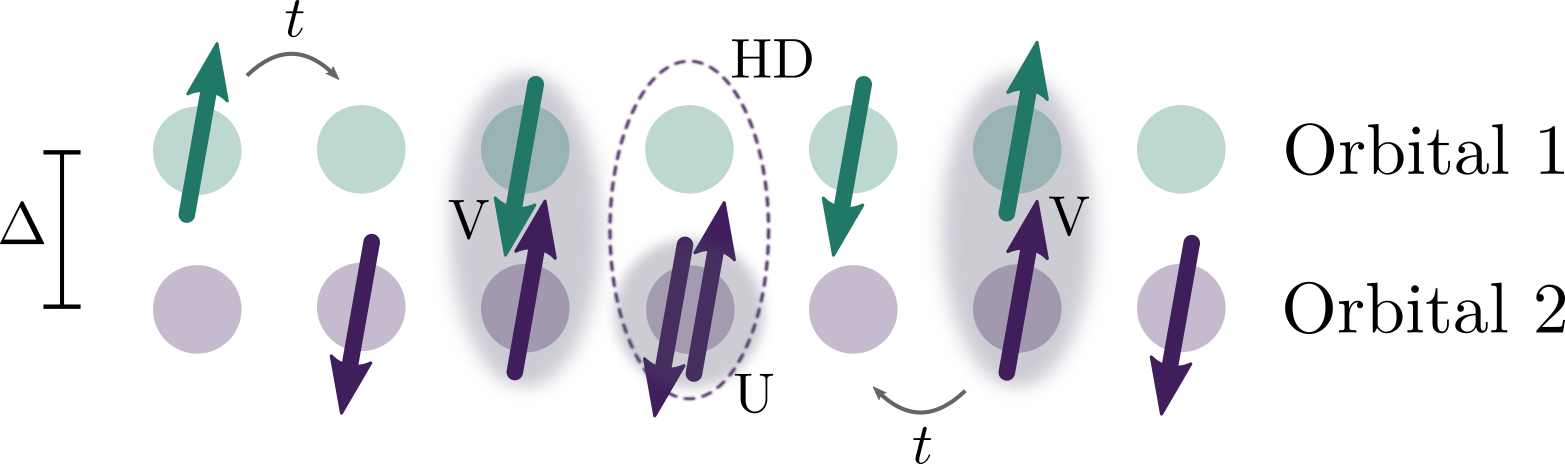}
  \caption{\label{int} Graphical representation of the model, Eq.~\ref{eq:KHM}. HD represents the local holon-doublon pair (marked using a dashed elliptical line, see text)}
 \end{figure}

$U$ ($V$) is the intraorbital (interorbital) Coulomb repulsion between electrons and we consider $U> V$ (see Fig. \ref{int}).
Here, $j$ denotes sites on an open chain, $\alpha=1,2$ and $\sigma$ are the orbital and spin indices, respectively.
The creation (destruction) operator on site $j$ is $c^{\dagger}_{j,\alpha\sigma}$ ($c_{j,\alpha\sigma}$); $n_{j,\alpha}=\sum_{\sigma}c_{j,\alpha\sigma}^{\dagger}c_{j,\alpha\sigma}$ is the on-site number operator for orbital $\alpha$ and $\Delta$ accounts for the crystal-field splitting between orbitals. 
A global gate voltage $V_g$ is introduced and subsequently the ground state is determined in Fock space; thereby the particle numbers $N_1$ and $N_2$ in each orbital are obtained fixing also the overall particle number $N=N_1+N_2$. Henceforth we set $\epsilon=-U/2-V$ in order to have a half-filled system when $V_g=0$ and $\Delta=0$.

The nearest-neighbor hopping is $t$ and no interorbital hybridization is included. We will consider $t=0.5$ and all energies are measured in units of $t$. 

We use the DMRG\cite{White,Karen1,Uli,Uli2,banuls,Karen2,ramasesha,KW} to calculate the ground state and its energy, the local density of states, the projections onto particular excitations and the optical conductivity, as described below.

\textit{Local density of states and projections}. The local density of states (LDOS) is obtained from the imaginary part of two dynamical response functions:
\begin{equation}
\begin{split}
A^{>}_{j,\alpha}(\omega)=-\frac{1}{\pi}\sum_{\sigma}\Im\langle c_{j,\alpha\sigma}(\omega+i\eta-H+E_{0})^{-1}c^{\dagger}_{j,\alpha\sigma}\rangle , \\
A^{<}_{j,\alpha}(\omega)=-\frac{1}{\pi}\sum_{\sigma}\Im\langle c_{j,\alpha\sigma}^{\dagger}(\omega+i\eta-H+E_{0})^{-1}c_{j,\alpha\sigma}\rangle ,
\end{split}
\label{eq:Aq<}
\end{equation}
where the expectation value is taken for the ground state of the system, energies are measured relative to the ground-state energy $E_0$, and the Lorentzian broadening is set to $\eta=0.1$.
The orbitally resolved local density of states follows as: $A_{LDOS,j,\alpha}\left(\omega\right) =  A^{>}_{j,\alpha}\left(\omega\right) + A^{<}_{j,\alpha}\left(-\omega\right)$.

The analysis of the LDOS is complemented by using projected response functions as previously defined\cite{nair,yurieldop,yurielhd}. In particular, we employ the projection onto HD local pairs (one orbital being doubly occupied and the other one empty) on site $j$: $A_{\text{HD},j,\alpha}(\omega)=A^>_{\text{HD},j,\alpha}(\omega)+A^<_{\text{HD},j,\alpha}(-\omega)$, where 

\begin{equation}
\begin{split}
A^>_{\text{HD},j,\alpha}\left(\omega\right) = -\frac{1}{\pi}\sum_{\sigma}\Im\braket{c_{j,\alpha\sigma}\left(\omega+i\eta-H+E_0\right)^{-1}X^\dagger_{\text{HD},j,\alpha}},\\
A^<_{\text{HD},j,\alpha}\left(\omega\right) = -\frac{1}{\pi}\sum_{\sigma}\Im\braket{c^\dagger_{j,\alpha\sigma}\left(\omega+i\eta-H+E_0\right)^{-1}X_{\text{HD},j,\alpha}},
\end{split}
\label{projector}
\end{equation}
with $X^\dagger_{\text{HD},j,\alpha}=P_{\text{HD},j,\alpha}c^{\dagger}_{j,\alpha\sigma}$, $X_{\text{HD},j,\alpha}=c_{j,\alpha\sigma}P_{\text{HD},j,\alpha}$ and $P_{\text{HD},j,1}=\ket{\uparrow\downarrow,0}\bra{\uparrow\downarrow,0}$ or $P_{\text{HD},j,2}=\ket{0,\uparrow\downarrow}\bra{0,\uparrow\downarrow}$. The bra and ket states represent the site configuration where the first (second) component corresponds to the electronic state in orbital 1 (2). \\

\textit{Optical conductivity}. Real part of the optical conductivity along the chain direction is obtained from\cite{tohyama,stephan}:
\begin{equation}
\Re \sigma\left(\omega\right) = -\frac{1}{L \pi\omega} \Im\braket{J^\dagger\left(\omega+i\eta-H+E_0\right)^{-1}J}
\label{conduct}
\end{equation}
where the current operator 
\begin{equation}
J=it\sum_{j,\alpha\sigma}\left( c^{\dagger}_{j+1,\alpha\sigma}c_{j,\alpha\sigma}-H.c.\right)
\label{current}
\end{equation}
is expressed in terms of the fermionic operators.

\section{Results}

In this section we present the DMRG results for the local density of states and for the optical conductivity. The particle numbers in both orbitals are fixed through the gate voltage $V_g$. Without loss of generality we consider $U=8$, $L=8$ physical sites (unless otherwise indicated), with two non-degenerate orbitals at each site and open boundary conditions. The LDOS is calculated near the center of the chain at site $j=4$ to minimize the boundary effects; we use up to $m=1024$ states for the ground-state energies (with a convergence error of $~10^{-8}$), $m=512$ for response functions and four to six sweeps. The value of $U$ was chosen to ensure a clear separation of the different subbands. However, the results are robust for a wide range of $U$ and $V$ as long as $U > V$. 

We study four different cases. Case A: An in-gap HD band in orbital 2 emerges with increasing $V$ when we start from a hole-doped orbital 1 and an insulating orbital 2. Case B: This HD band is shown to exist when orbital 1 turns doped with decreasing $V_g$. Case C: The formation of an HD metal upon changing the gate voltage.
And finally case D: The presence of an HD band in orbital 1 is verified also when orbital 2 is electron doped. \\

\subsubsection{Local density of states: Emergence of an HD band}
\begin{figure}[h!]
\centering
\begin{subfigure}{0.48\textwidth}
    \includegraphics[width=1\textwidth]{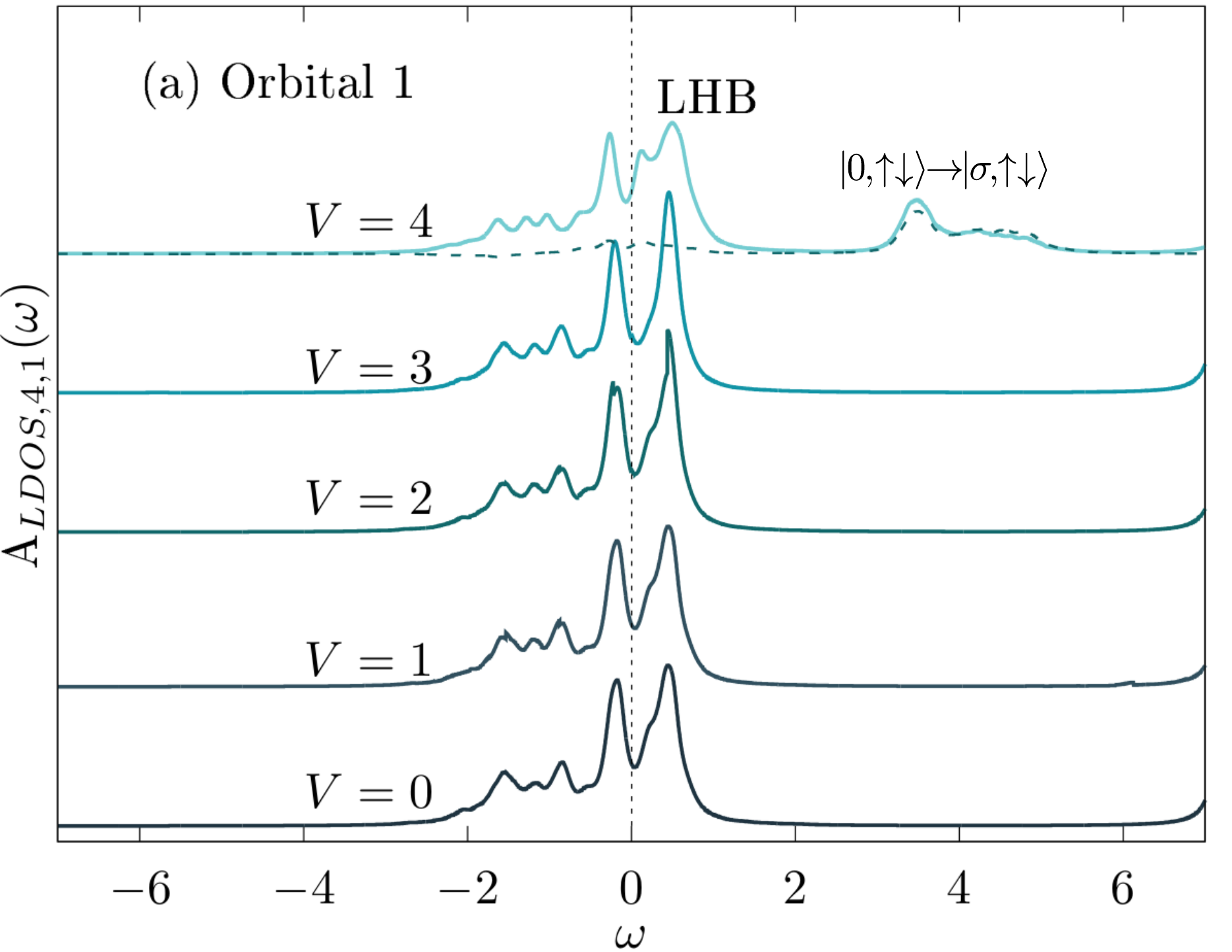}
\end{subfigure}

\begin{subfigure}{0.48\textwidth}
    \includegraphics[width=1\textwidth]{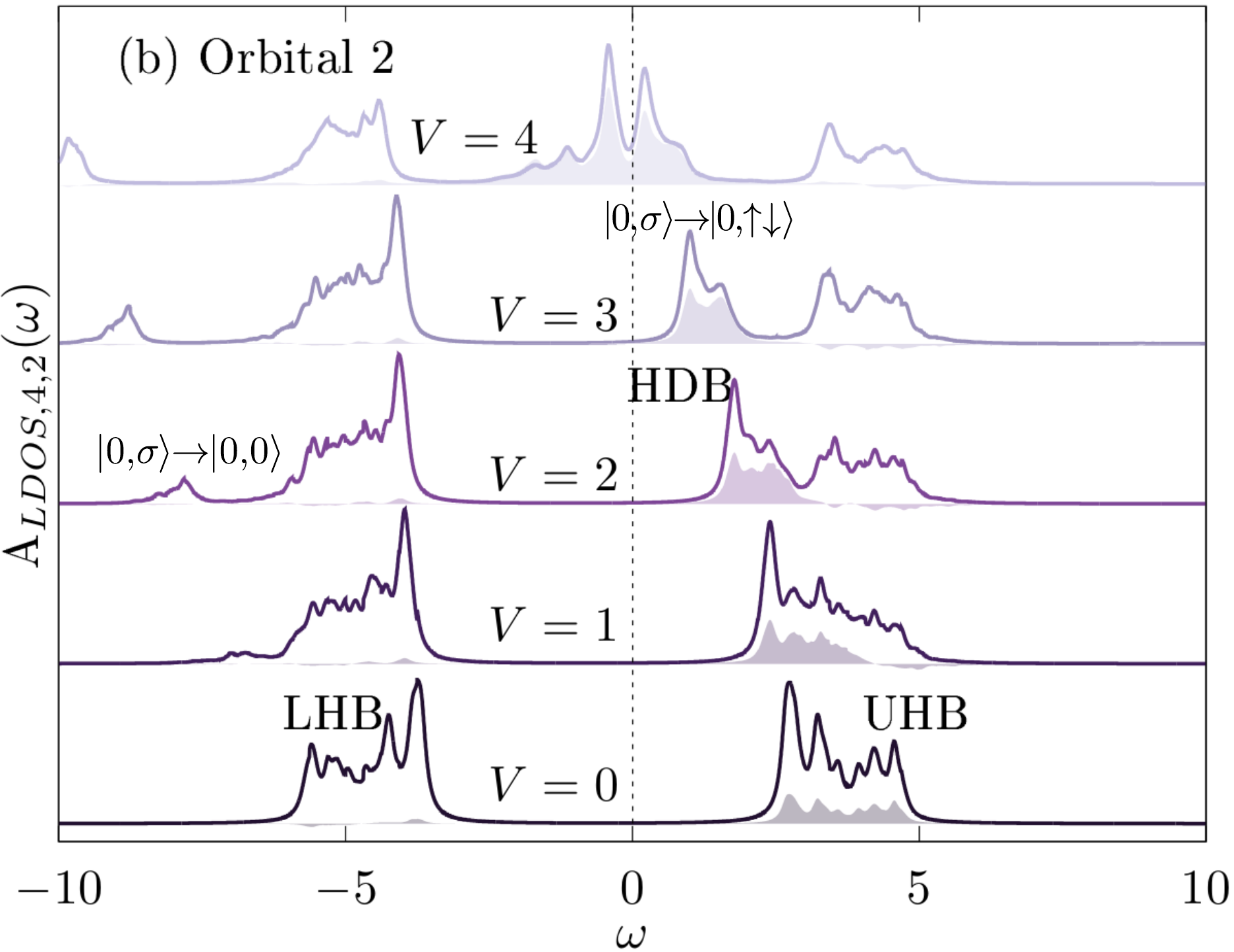}
\end{subfigure}
\caption{Case A: LDOS for a sequence of $V$ and $L=8$, $U=8$, $\Delta=4$ and $V_g=-3.5$. Starting from a hole-doped orbital 1 and an insulating half-filled orbital 2 $V$ is increased (from bottom to top). A finite $V$ generates a band of HD excitations in orbital 2 which crosses $\omega=0$ for $V=4$. The shaded regions correspond to the projected HD spectra (Eqs.~\ref{projector}), and the dashed line in orbital 1 indicates projections on local excited configurations of the $|\sigma, \uparrow\downarrow\rangle$ type ($\sigma$ represents the spin degree of freedom) following Ref. \cite{nair}. The LHB and the UHB of orbital 2 are marked, as well as the LHB of orbital 1. The UHB of orbital 1 is not shown (out of scale, around $\omega =8$). Note the difference in the $\omega$ scales between (a) and (b)} 
\label{vsV}
\end{figure}

\textit{Case A}: First we consider the effect of changing the interorbital interaction $V$ starting from a situation in which orbital 1 is hole doped and orbital 2 is a Mott insulator (half-filled) with $V=0$.

In all the results shown in Fig. \ref{vsV}, orbital 1 is quarter-filled ($N_1=4$) so it is doped with holes. For $V=0$ we immediately identify the lower (LHB) and upper (UHB) Hubbard bands in orbital 2. Upon turning on the interorbital Coulomb repulsion, spectral weight is transferred from the UHB towards lower energies. For $V=2$ a new separated subband in the gap between the Mott-Hubbard bands has formed. This situation is similar to the already reported results for degenerate orbitals\cite{nair,yurieldop,yurielhd}.

To characterize this in-gap band, we have calculated the projected LDOS onto HD states using Eqs.~\ref{projector}. These are represented by the shaded areas in Fig.~\ref{vsV}b for orbital 2 and indicate that there is a large weight which corresponds to HD excitations, supporting the notion of an in-gap HD band (HDB). In the local or atomic limit representation, this band corresponds to excitations of the type $\ket{0,\sigma}\rightarrow\ket{0,\uparrow\downarrow}$ (following the notation given below Eqs.~\ref{projector}, where  $\sigma$ represents a generic spin). A simple calculation in the atomic limit identifies the energy for this excitation $\omega =U/2-V-\Delta-V_g=3.5-V$ for the parameters used here, which coincides approximately with the mean energy of the HD band for all values of $V$.
When $V\geq 1$ there is a structure located at energy $\omega = -U/2 -V - V_g - \Delta = -4.5 -V$ that corresponds to transitions of the type $\ket{0,\sigma} \rightarrow \ket{0,0}$, i. e. creating an empty site by removing an electron from orbital 2.

For $V=4$ the HDB in orbital 2 crosses $\omega=0$ indicating a metallic behavior with $N_2=9$ which is one electron above half-filling. In this case orbital 2 is no longer a Mott insulator but rather electron doped, and the LDOS in orbital 1 is substantially modified with new spectral weight appearing at higher energies around $\omega = 4$. This new structure corresponds to excitations of the type indicated in brackets in Fig.~\ref{vsV}a: $\ket{0,\uparrow\downarrow}\rightarrow\ket{\sigma,\uparrow\downarrow}$ as defined in Ref. \cite{nair}.
 The atomic limit calculation leads to an energy difference between those two states of $\omega =V-U/2-V_g=3.5$ for the parameters considered here, which is close to the energy of this structure.
This means that HD configurations with the hole in orbital 1 and the doublon in orbital 2 are present in the ground state and that creating a particle in orbital 1 implies a particle to a HD configuration.

The LDOS for orbital 2 shows a clear separation between the UHB and the HDB of approximately $V$.

\begin{figure}[h!]
\centering
\begin{subfigure}{0.48\textwidth}
    \includegraphics[width=1\textwidth]{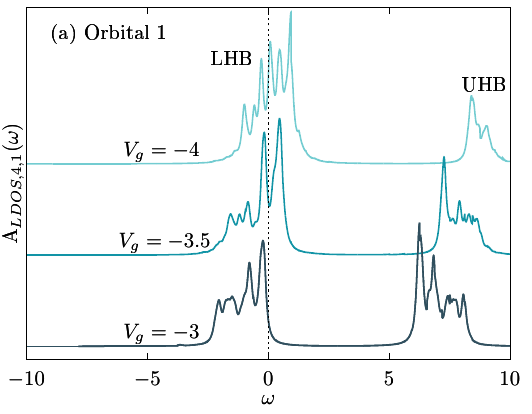}
\end{subfigure}

\begin{subfigure}{0.48\textwidth}
    \includegraphics[width=1\textwidth]{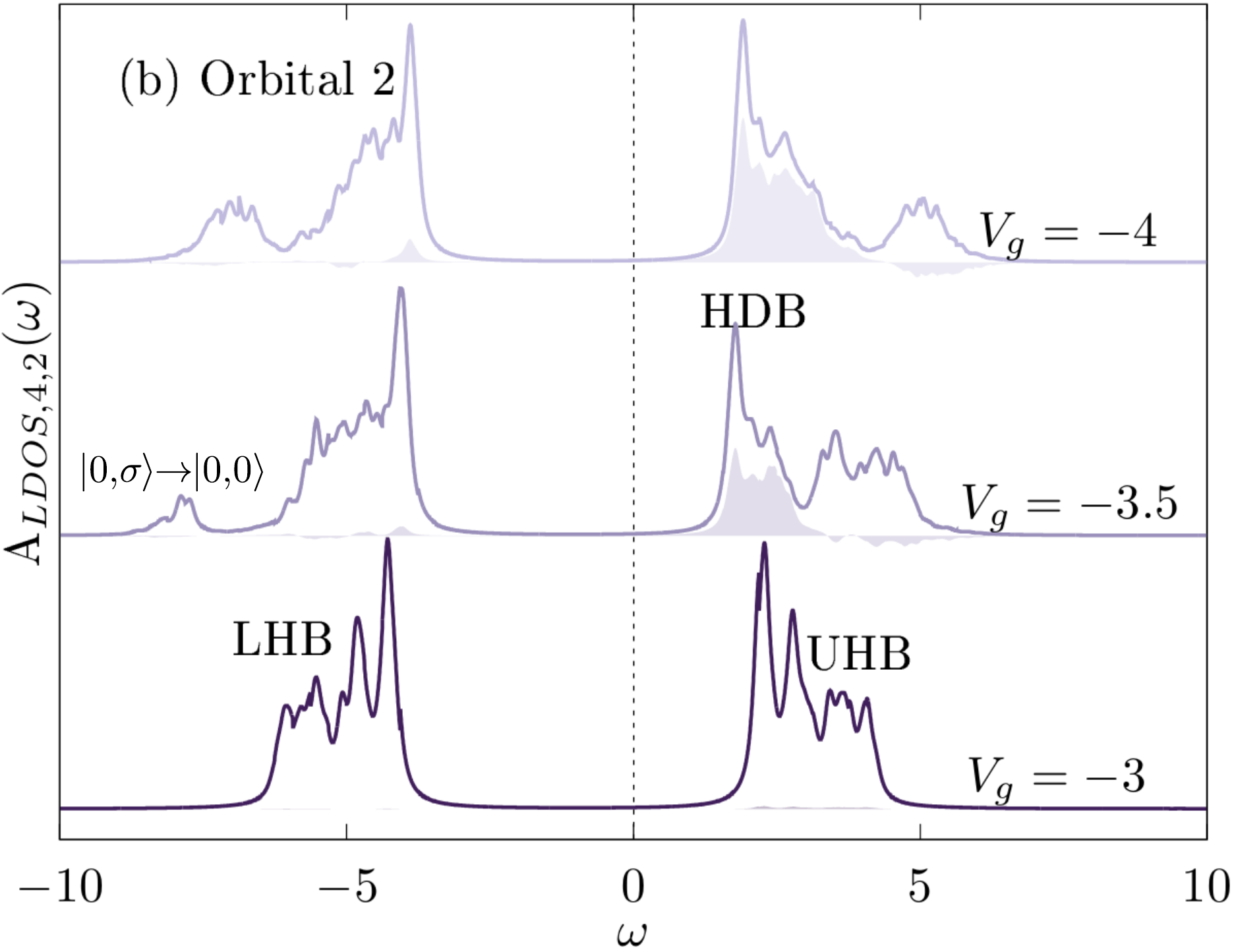} 
\end{subfigure}
\caption{Case B: LDOS for a sequence of $V_g$ and $L=8$, $U=8$, $V=2$, $\Delta=4$. Starting from both orbitals at half-filling and doping holes into orbital 1 (bottom up). As soon as holes enter orbital 1 a HDB forms in orbital 2.
 The shaded regions correspond to the projected HD excitations (see Eqs.~\ref{projector})}
\label{vsmuV2}
\end{figure}

\textit{Case B:} In Fig.~\ref{vsmuV2} we analyze the behavior of the LDOS starting from $V_g=-3$ with both orbitals in a half-filled Mott insulating state and lower $V_g$ for fixed $V=2$. 
The tail crossing $\omega=0$ for orbital 1 at $V_g=-3$ is due to the Lorentzian broadening.
By lowering $V_g$ orbital 1 becomes hole doped (with $N_1=6$ and $N_1=4$ for $V_g=-3.5$ and $V_g=-4$, respectively), 
the UHB structure for $\omega>0$ in orbital 2 is modified by transferring weight to a new low-energy band which has a large component of HD-excitations. As these excitations are formed by a hole in orbital 1 and a doubly occupied state in orbital 2, they can only exist in this large $U$ case when the particle number in orbital 1 shrinks below half filling. The new structure below the LHB in orbital 2 corresponds to the transition $|0,\sigma\rangle \rightarrow |0,0\rangle$.

\begin{figure}[h!]
\centering
\begin{subfigure}{0.48\textwidth}
    \includegraphics[width=1\textwidth]{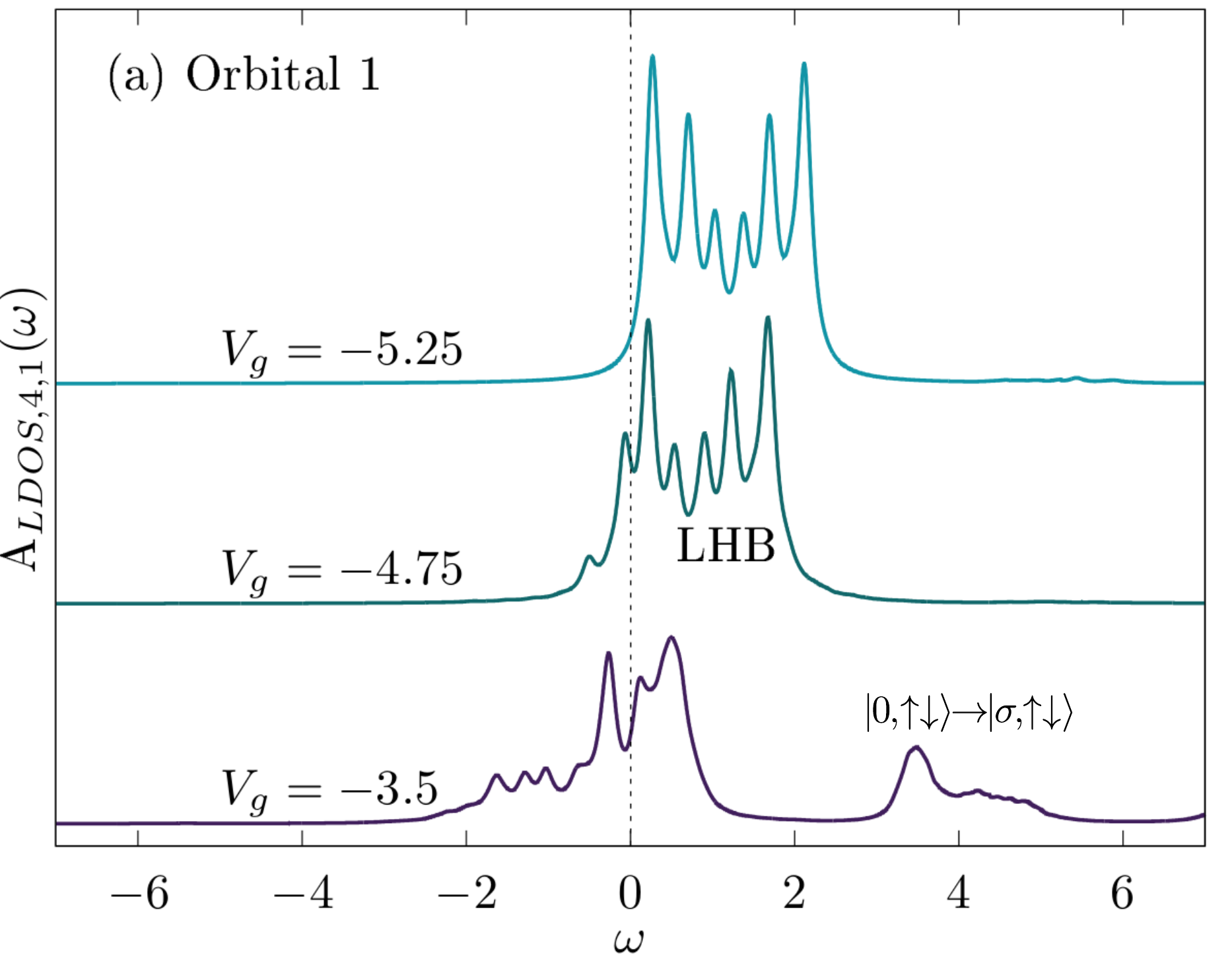}
\end{subfigure}

\begin{subfigure}{0.48\textwidth}
    \includegraphics[width=1\textwidth]{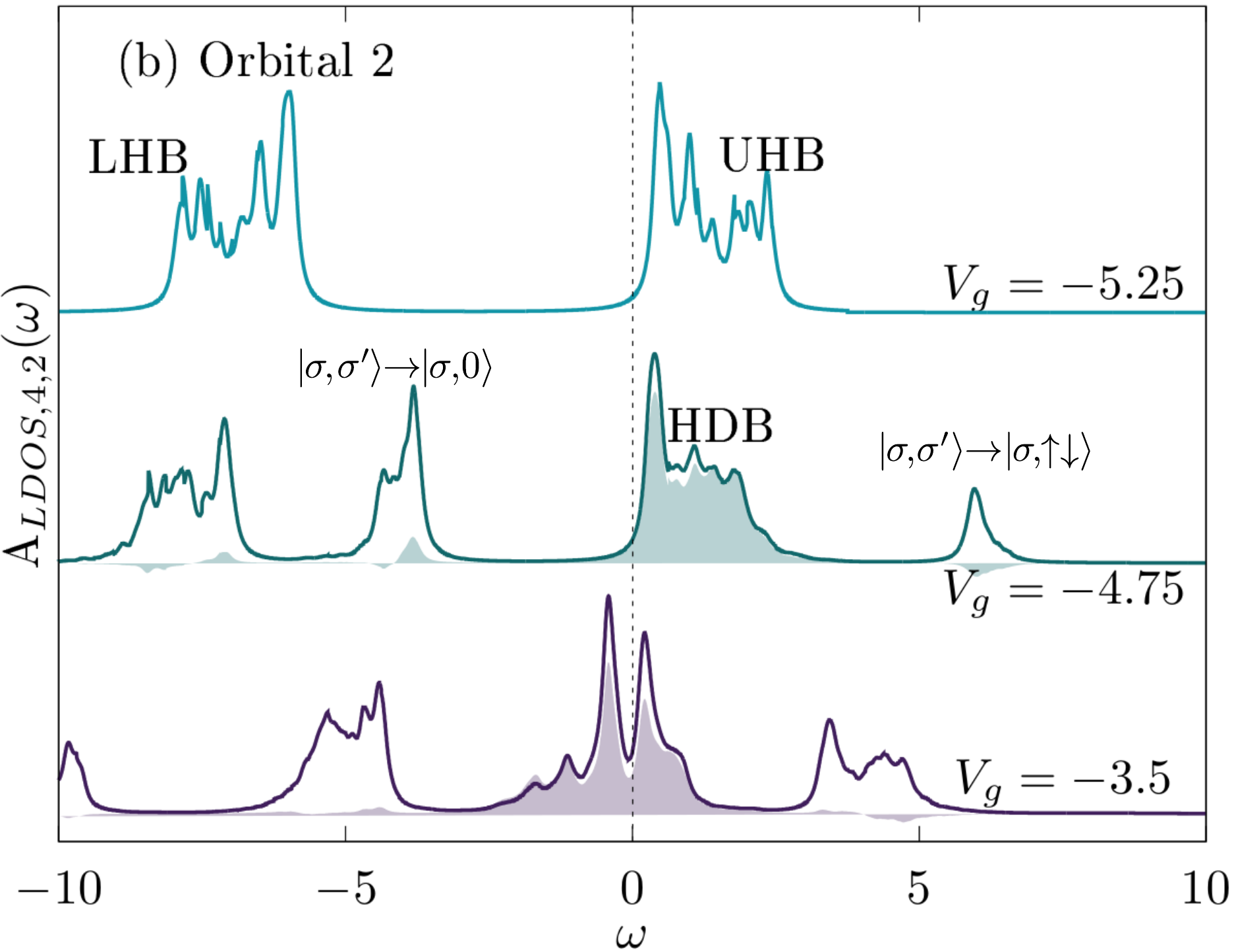} 
\end{subfigure}
\caption{Case C: LDOS for a sequence of $V_g$ for $L=8$, $U=8$, $V=4$ and $\Delta=4$. Starting from a hole-doped orbital 1 and an electron-doped orbital 2, $V_g$ is lowered (bottom up)\cite{clarification}. HD excitations with doublons residing in orbital 2 are shaded using the projection in Eqs. \ref{projector}. Note the difference in the $\omega$ scales between (a) and (b)}
\label{vsmuV4}
\end{figure}

\textit{Case C:} In Fig.~\ref{vsmuV4} we show the LDOS starting from both orbitals away from half-filling and lowering $V_g$ again. In this case, orbital 1 is quarter filled (hole doped) and orbital 2 is electron doped ($N_2=10$) at $V_g=-3.5$ due to the increased interorbital interaction $V=4$ (with respect to Case B), with the HDB located around $\omega=0$, thus indicating a  metallic character. Doublons are naturally formed in the electron-doped orbital 2 and pair up with a hole in orbital 1.
Upon lowering $V_g$ orbital 2 returns to a half-filled insulator and also empties orbital 1  ($N_1=2$ and $N_1=0$ for $V_g=-4.75$ and $V_g=-5.25$, respectively). As long as orbital 1 is not empty, the structure around $\omega= 6$ in orbital 2,
corresponds mainly to transitions of the type $|\sigma , \sigma'\rangle \rightarrow |\sigma,\uparrow\downarrow\rangle$ (indicated in Fig.~\ref{vsmuV4}b) whose energy in the atomic limit is $\omega=U/2-V_g-\Delta=4.75$. 
In addition, the structure around $\omega=-4$ corresponds mainly to transitions $|\sigma , \sigma'\rangle \rightarrow |\sigma,0\rangle$ whose energy in the atomic limit is $\omega=-U/2-V_g-\Delta=-3.75$.

\begin{figure}[h!]
\centering
\includegraphics[width=0.48\textwidth]{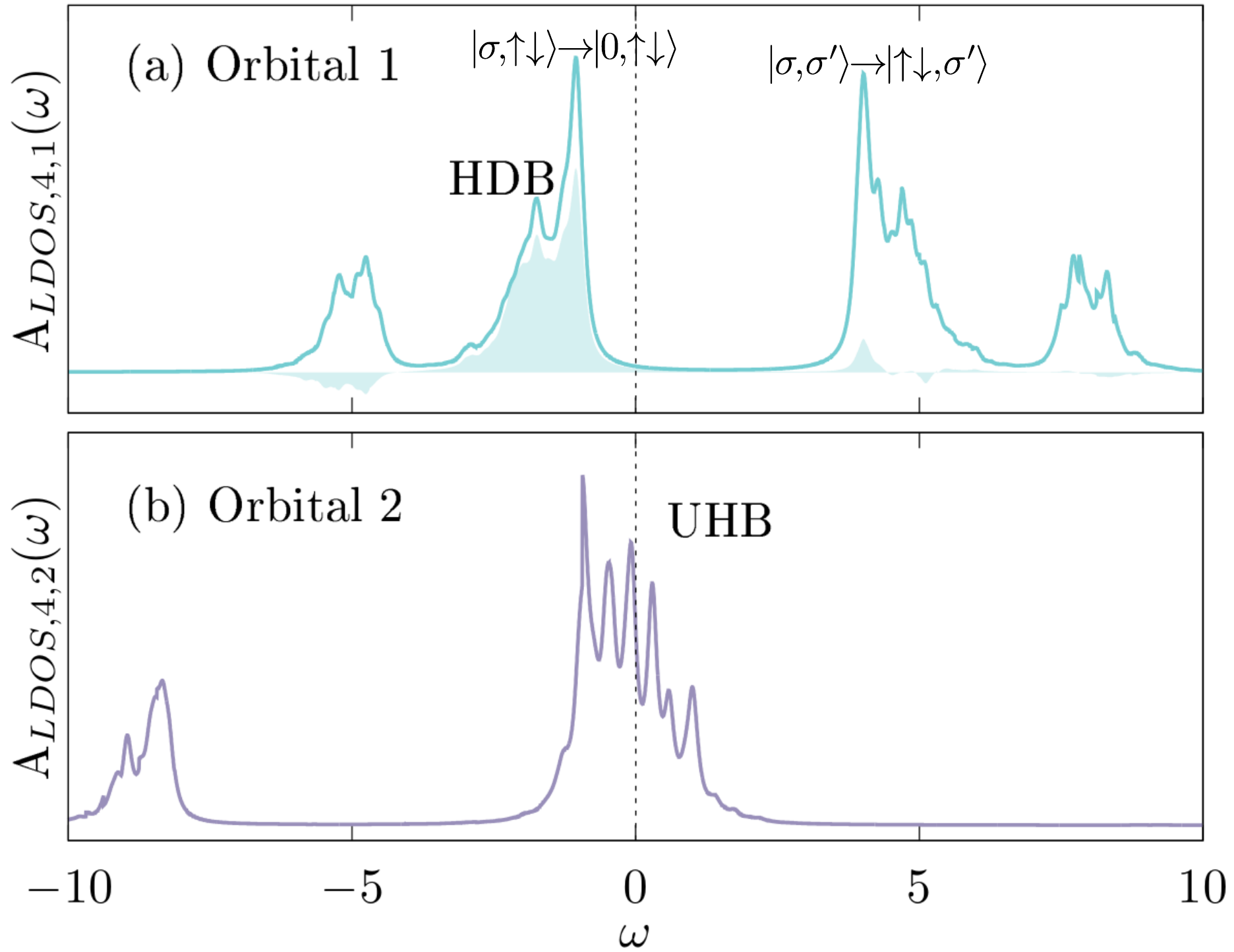}  
\caption{Case D: LDOS for $L=8$, $U=8$, $V=3$, $\Delta=4$ and $V_g=0$: a half-filled orbital 1 and an electron-doped orbital 2 enable the existence of HD-pairs  when removing a particle in orbital 1. The process is pictured as: $\ket{\sigma, \uparrow\downarrow}\rightarrow\ket{0,\uparrow\downarrow}$. 
The transitions of type: $\ket{\sigma, \sigma'} \rightarrow \ket{\uparrow\downarrow, \sigma'}$ are associated to the structure around $\omega\sim 4$}
\label{edoping}
\end{figure}

\textit{Case D:} A similar situation occurs for $\Delta=4$ and $V_g=0$, when orbital 1 is in a Mott insulating half-filled state and orbital 2 is three quarter filled (electron doped).
Because of the overlap between HD states and the ground state of the system after removing a particle from orbital 1 (when it is singly occupied and when orbital 2 is doubly occupied) a HDB appears in Fig. \ref{edoping}a. The shaded region indicates this overlap with local HD pairs.
The latter transition in the atomic limit,  $\ket{\sigma , \uparrow\downarrow} \rightarrow \ket{0,\uparrow\downarrow}$, has an energy $\omega=-U/2+V-V_g=-1$.
Instead, the structure at $\omega \sim 4$ corresponds to the transitions $|\sigma ,\sigma '\rangle \rightarrow |\uparrow\downarrow, \sigma '\rangle$ whose energy in the atomic limit is $\omega=U/2-V_g=4$.

\subsubsection{Optical Conductivity}

\begin{figure}[h!]
\centering
\includegraphics[width=0.5\textwidth]{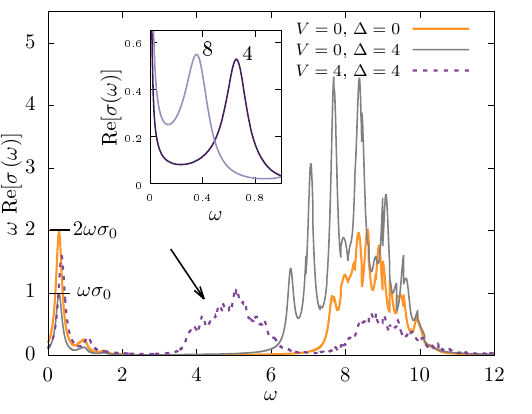}
\caption{ Optical conductivity multiplied by $\omega$ for $L=8$, $U=8$ and $V_g=-3.5$. The orange line shows the case $V=\Delta=0$ where both orbitals are hole doped and contribute in the same way to the conductivity; the grey line corresponds to the case $V=0$, $\Delta=4$, in which orbital 1 is hole doped and orbital 2 is a Mott insulator (see Fig.~\ref{vsV}), and the purple dashed line shows the result when both orbitals are metallic but the metallicity of orbital 2 stems from a HDB ($V=\Delta=4$).
The arrow indicates the absorption band due to the HDB.
 The inset shows the conductivity Re$\left[\sigma(\omega)\right]$ for $L=4$ and $L=8$ for this latter case}
\label{optical}
\end{figure}

In Fig.~\ref{optical} we show results for the optical conductivity. The main panel displays $\omega $Re$\left[\sigma(\omega)\right]$ and thereby avoids the $\omega\rightarrow 0$ rise of Re$\left[\sigma(\omega)\right]$ which is solely due to Lorentzian broadening. Eq.~\ref{conduct} was employed for $L=8$, $U=8$, $V_g=-3.5$ and three qualitatively different cases:
i) two independent degenerate orbitals ($V=\Delta=0$) i.e. two uncoupled hole-doped Hubbard chains with $N_1=N_2=6$ particles; ii) two independent non-degenerate orbitals where orbital 1 is hole doped and orbital 2 is a half-filled Mott insulator ($V=0$, $\Delta=4$); and the case with a HDB crossing $\omega=0$ in orbital 2 as in Fig.~\ref{vsV} for $V=\Delta=4$).

The metallic character of the whole system with open boundary conditions is reflected in the finite-size scaling behavior of the peak structure close to $\omega \sim 0$. This peak evolves into the zero-frequency Drude peak when the system size is continuously increased\cite{shastry}. 
The scaling for the lowest-frequency peak in Re$\left[\sigma(\omega)\right]$ is evaluated in 
the inset in Fig.~\ref{optical}. Here the optical conductivity is shown for $L=4$ and $L=8$ for constant electronic density.

When comparing the three cases in the main panel, distinctly different metallic characteristics are observed: 
for $V=\Delta=0$ both independent hole-doped orbitals contribute equally to the optical conductivity and the Drude peak weight is twice as large as the conductivity $\sigma_0$ of the case $V=0$ and $\Delta=4$ for which one of the orbitals is quarter filled and the other one half filled. Instead, for $V=\Delta=4$, where both orbitals are doped and a HDB exists around $\omega=0$ (Fig.~\ref{vsV}), 
 the weight of the Drude peak is significantly reduced compared to the two independent metallic orbitals, indicating the formation of a correlated metal. In other words, when both orbitals are metallic but the metallicity in one of them (orbital 2 in this case) is due to the presence of a HDB at $\omega=0$, the conductivity is reduced as compared to the case in which both orbitals are independent and metallic.

This figure also reveals how the absorption band is modified by the presence of $V$. For $\Delta=4$ and $V=0$ there is a conspicuous incoherent structure centered at $\omega\cong 8$ which corresponds to the well-known transitions involving the UHB. However, when $V=4$, a new absorption band appears at lower energy, around $\omega \cong 5$, in addition to the one involving the UHB. This band is entirely determined by the HD excitations which form a novel structure within the Hubbard bands (shaded region in Fig.~\ref{vsV}b). The higher-energy absorption corresponds to excitations involving the remainder of the UHB containing excitations not participating in the HDB formation.

\newpage
\section{Conclusions}
Our DMRG study reveals the emergence of a band within the Mott-Hubbard gap as a consequence of a finite interorbital Coulomb repulsion $V$. This structure
maintains a large weight when projected onto holon-doublon excitations, which we call the holon-doublon band (HDB). Although a similar structure was already previously reported\cite{nair}, here, for non-degenerate orbitals, we demonstrate that an insulating orbital can become gapless, if the other orbital provides mobile carriers even in the absence of inter-orbital hopping.
When $V$ is large enough the HDB crosses $\omega=0$ and the system becomes an unconventional correlated metal.
The optical conductivity differentiates between the distinct metallic behaviors. When both orbitals are metallic and $V=0$ (independent orbitals), 
they both contribute equally to the conductivity.
However, when one of the orbitals is doped and $V$ is large enough, the other orbital can turn gapless
due to the existence of the HDB. However, the conductivity of this correlated HD metal is much lower than that for the previous case, indicating the existence of unusual and heavier current-carrying quasiparticles with holon-doublon character. In this case the absorption spectrum in the optical conductivity is dominated by this HDB and at lower energies it precedes the absorption due to the upper Hubbard band.

The identification of novel bands within the electronic structure of a prominent interacting electron model holds the potential to unveil previously unexplored structure in photoemission and absorption spectra in correlated materials.

\section{Acknowledgments}
NAB and KH acknowledge support from ICTP through the STEP and Associates Programmes respectively, and from the PICT 2018-01546 grant of the Agencia I+D+i, Argentina.


\begin{references}
\bibitem{clarification} The slight asymmetry between the LHB and UHB for $V_g=-5.25$ is due to small uncertainties in the DMRG which is less precise for higher energies.
\end{references}
\end{document}